\documentstyle[aps,prb,multicol]{revtex}
\begin{document}
\draft 
\date{\today}
\title{Proximity Effect Enhancement Induced by Roughness of SN
Interface}
\author{R.G.~Mints, and I.B.~Snapiro}
\address{School of Physics and Astronomy, Raymond and Beverly Sackler
Faculty of Exact Sciences,\\ 
Tel Aviv University, Tel Aviv 69978, Israel}
\maketitle
\begin{abstract} 
Critical temperature reduction $\Delta T_c$ is considered for a thin
film of a layered superconductor (S) with a rough surface covered by a
thick layer of a normal metal (N). The roughness of the SN interface
increases the penetration of electrons from the normal metal into the
superconductor and leads to an enhancement of the proximity effect. The
value of $\Delta T_c$ induced by the roughness of the SN interface can
be much higher than $\Delta T_c$ for a film with a plain surface for an
extremely anisotropic layered superconductor with the coherence lengths
$\xi_a,\xi_b\gg\xi_c$.
\end{abstract}
\begin{multicols}{2}
Proximity effects are determined by the penetration of electrons from
the normal metal into the superconductor and by the penetration of
Cooper pairs from the superconductor into the normal metal.\cite{r1,r2} A
dramatic manifestation of proximity effects is the reduction of the
critical temperature $\Delta T_c$ of a thin superconducting film covered
by a thick layer of a normal metal.\cite{r1} In the case of an
isotropic superconductor with a plain SN interface the value of $\Delta
T_c$ within the Ginzburg-Landau approach is given by the formula\cite{r3}
\begin{equation} 
{\Delta T_c\over T_c} = -{\gamma^2\pi^2\over 4}{\xi^2(0)\over d^2}, 
\label{eq1} 
\end{equation} 
where $\gamma\approx 0.74$ is a numerical factor, $\xi (0)$ is the
coherence length at zero temperature, $d$ is the thickness of the
superconducting film ($d\gg\xi(0))$. 
\par 

Proximity effects in anisotropic high-temperature copper oxide
superconductors are currently under thorough experimental and
theoretical study.\cite{r4} The reduction of the transition temperature
for an anisotropic superconductor depends on the orientation of the SN
interface relative to the symmetry axes of the superconductor. The value
of $|\Delta T_c|$ is maximal when the SN interface is perpendicular to
the axis corresponding to the maximum value of the coherence length. For
an anisotropic superconductor with a rough SN interface the local
orientation of the film surface relative to the symmetry axes is varying
along the surface. A certain average value of $\xi^2(0)$ appears then in
Eq.~(\ref{eq1}). Depending on the roughness this average value can be
determined by the highest of the coherence lengths even if for the same
orientation of a plain surface $\Delta T_c$ is determined by the
smallest of the coherence lengths. Therefore, for a strongly anisotropic
superconductor the roughness of the SN interface can drastically
increase the proximity effect. 
\par

In the present paper we treat the reduction of the critical temperature
$\Delta T_c$ for a rough superconducting film of an anisotropic
superconductor covered by a thick layer of a normal metal and located on
top of a substrate from an insulator with a plane surface ($z=0$), where
the $z$ axis is along the $c$ direction and the $xy$ plane is parallel
to the $ab$ planes. The average thickness of the film $d$ is considered
to be bigger than the coherence lengths at zero temperature
$\xi_{\mu}(0)$ ($\mu =a,b,c$), {\it i.e.}, $d\gg\xi_\mu (0)$. We assume
that the roughness is small and describe the SN interface as
$z=z_S(x,y)=d+f(x,y)$, with $|f(x,y)|\ll d$ and a zero average value
$\langle f\rangle$ of $f(x,y)$
\begin{equation}
\langle f\rangle={1\over A}\int dxdy\, f(x,y)=0,
\label{eq2}
\end{equation}
where A is the area of the film. The typical length scale $l$ of
variations of the film thickness $f(x,y)$ is considered to be from the
interval $\xi_{\mu}(0)\ll l\ll\xi_{\mu}(T)$. 
\par

To find the reduction of the transition temperature we treat the
Ginzburg-Landau free energy 
\begin{equation} 
{\cal F}={H_c^2\over 4\pi}
\int d^3{\bf r}\,\Bigl [-\Psi^2 +{1\over2}\,\Psi^4 + 
\xi_{\mu}^2\Big |{{d\Psi}\over {dx_{\mu}}}\Big |^2\Bigr], 
\label{eq3} 
\end{equation} 
where the order parameter $\Psi$ depends on the three coordinates $x$,
$y$, and $z$ in the case of a rough SN interface and $H_c$ is the
critical magnetic field. To account for this dependence and to find the
value of ${\cal F}$ we use a trail function in the form
\begin{equation} 
\Psi=\Phi (x,y)\cos{\pi z\over 2[d+f(x,y)]}.
\label{eq4}
\end{equation}
This function satisfies to the standard Ginzburg-Landau boundary
conditions at the film surfaces\cite{r3}
\begin{equation}
\Psi\Big |_{z=d+f(x,y)} =0,\quad 
{\partial\Psi\over {\partial z}}\Big |_{z=0} =0
\label{eq5}
\end{equation}
and to find $\Phi (x,y)$ we have to minimize the free energy. 
\par

Substituting Eq.~(\ref{eq4}) into the functional (\ref{eq3}) and
assuming that $|f|\ll d$ we obtain after integration over $z$
\begin{eqnarray}
{\cal F}&=&{H_c^2\over 8\pi}\,\int dxdy\,
\Big\{\Big [\pi^2\xi_c^2 +{\pi^2+3\over 3}\,\xi_i^2f_i^2-4d^2\Big ]\,
{\Phi^2\over 4d}+
\nonumber\\
& &d\xi_i^2\Phi_i^2 +\xi_i^2f_i\Phi\Phi_i +{3d\over 8}\,\Phi^4\Big\}
\label{eq6}
\end{eqnarray}
where $i=a,b$, $f_i=\partial f/\partial x_i$, and 
$\Phi_i=\partial \Phi/\partial x_i$.
Minimization of the free energy (\ref{eq6}) results in the following 
equation for $\Phi(x,y)$
\begin{eqnarray}
& &\xi_i^2\Phi_{ii} +{\xi_i^2f_i\over 2d}\,\Phi_{i}+\nonumber\\ 
& &\Phi\Big [1-{\pi^2\xi_c^2\over 4d^2} -{(3+\pi^2)\,\xi_i^2f_i^2\over 12d^2} +
{\xi_i^2f_{ii}\over 2d}\Big ] -{3\over 4}\, \Phi^3=0 
\label{eq7}
\end{eqnarray}
\par

We are interested in a solution of equation (\ref{eq7}) imposed by the
roughness of the SN interface. This solution has a length scale of the
order of the length scale $l$ of variations of the film thickness
$f(x,y)$. We take $\Phi (x,y)$ in the form
\begin{equation}
\Phi =\Phi_0 +\Phi_1(x,y),
\label{eq8}
\end{equation}
where $\Phi_0=Const$, $|\Phi_1|\ll\Phi_0$, and length scale of
$\Phi_1(x,y)$ is of the order of $l$. It follows then from
Eq.~(\ref{eq7}) that with the accuracy of $|f(x,y)|/d\ll 1$ we have
\begin{equation}
\Phi_1=-\Phi_0\,{f\over 2d}.
\label{eq9}
\end{equation}
\par

Finally, using Eqs.~(\ref{eq6}), (\ref{eq8}), and (\ref{eq9}) we obtain
for the Ginzburg-Landau free energy
\begin{equation}
{\cal F}= {\cal F}_0\,
\Big\{\Big [{6+\pi^2\over 12}{\xi_i^2\langle f_i^2\rangle\over d^2} 
+{\pi^2\over 4}{\xi_c^2\over d^2} -1\Big ]\Phi_0^2
+ {3\over 8}\,\Phi_0^4\Big\},
\label{eq10}
\end{equation}
where ${\cal F}_0=AdH_c^2/128\pi$. The temperature of the 
superconducting-to-normal transition is determined by the change of sign of the
coefficient at $\Phi_0^2$ in the equation (\ref{eq10}) for the free
energy. Taking the temperature dependence of the coherence lengths near
$T_c$ as 
\begin{equation} 
\xi_\mu =\gamma\xi_\mu(0)(1-T/T_c)^{-1/2} 
\label{eq11} 
\end{equation} 
we obtain the reduction of the transition temperature $\Delta T_c$ in the 
following final form
\begin{equation} 
{\Delta T_c\over T_c}=-\gamma^2\,{\pi^2\over 4}\,\Bigl [{\xi_c^2(0)\over d^2} +
{6+\pi^2\over 3\pi^2}{\xi_i^2(0)\langle f_i^2\rangle\over d^2}\Bigr ].
\label{eq12} 
\end{equation} 
\par 
In the case of a plain SN interface the slopes $f_i=0$ and
Eq.~(\ref{eq12}) coincides with Eq.~(\ref{eq1}). For a rough SN
interface there is an enhancement of the proximity effects, the higher
is the anisotropy the higher is this enhancement. A considerable
enhancement of the reduction of the critical temperature appears for an
extremely anisotropic layered superconductor with $\xi_{a,b}\gg\xi_c$
even if the roughness is relatively moderate $\langle f_i^2\rangle\sim
1$. Assuming, for simplicity, that $\xi_a=\xi_b=\xi_{\rm ab}$ and
$\langle f_a^2\rangle =\langle f_b^2\rangle =\langle f'^2\rangle$ we
derive from Eq.~(\ref{eq12}) a simple formula to estimate the effect
of the roughness of the SN interface on $\Delta T_c$, namely, 
\begin{equation}
{\Delta T_c\over T_c}=-1.35\,\Bigl [{\xi_c^2(0)\over d^2} +
1.07\,\langle f'^2\rangle\,{\xi_{\rm ab}^2(0)\over d^2}\Bigr ].
\label{eq13} 
\end{equation} 
\par

In conclusion, we have studied the reduction of the critical temperature
for a thin rough film of a layered superconductor covered by a thick
layer of a normal metal. We have found that in the case of an extremely
anisotropic layered superconductor the roughness of the SN interface can
significantly increase the reduction of the critical temperature
compared with the case of a plain film.
\par

We acknowledge the support of the German-Israeli Foundation for Research
and Development, Grant \mbox{\# 1-300-101.07/93}. RGM is grateful to
G.~Deutscher and E.~Polturak for stimulating and informative
discussions. 
\par

\end{multicols}
\end{document}